\documentclass[12pt, letterpaper]{article}

\usepackage{amssymb}
\usepackage{amsmath}
\usepackage{verbatim}
\usepackage{color}

\mathversion{normal}

\makeatletter \@addtoreset{equation}{section} \makeatother

\newcommand{\bea}{\begin{eqnarray}}
\newcommand{\eea}{\end{eqnarray}}
\newcommand{\be}{\begin{eqnarray}}
\newcommand{\ee}{\end{eqnarray}}
\newcommand{\ben}{\begin{eqnarray*}}
\newcommand{\een}{\end{eqnarray*}}
\newcommand{\beq}{\begin{equation}}
\newcommand{\eeq}{\end{equation}}

\ifx\genfrac\sdflkaj

\else

\fi
\newcommand{\sfrac}[2]{{\textstyle\frac{#1}{#2}}}
\ifx\half\sdflkaj
\newcommand{\half}{\sfrac{1}{2}}
\fi

\let\oldPhi=\Phi
\let\oldPsi=\Psi
\renewcommand{\Phi}{\mathnormal{\oldPhi}}
\renewcommand{\Psi}{\mathnormal{\oldPsi}}

\newenvironment{myeqnarray}{\arraycolsep0pt\begin{eqnarray}}{\end{eqnarray}\ignorespacesafterend}
\newenvironment{myeqnarray*}{\arraycolsep0pt\begin{eqnarray*}}{\end{eqnarray*}\ignorespacesafterend}

\def\[{\begin{equation}}
\def\]{\end{equation}}
\def\<{\begin{myeqnarray}}
\def\>{\end{myeqnarray}}

\ifx\href\asklfhas\newcommand{\href}[2]{#2}\fi

\begin{document}

\begin{titlepage}
\vspace{1in}
\begin{flushright}
\end{flushright}
\vspace{1cm}

\begin{center}
{\Large\bf Spectral curve for $\gamma$-deformed $AdS/CFT$}\\
\vskip 12mm

{Minkyoo Kim}
\vskip 1cm
{\small \it MTA Lend\"ulet Holographic QFT Group,\\
 Wigner Research Centre for Physics,\\
H-1525 Budapest 114, P.O.B. 49, Hungary}\\
\vskip3mm
\small{\href{mailto:minkyoo.kim@wigner.mta.hu}{\tt minkyoo.kim@wigner.mta.hu}}

\vspace{.6in}
\vskip 0.6cm

\end{center}
\begin{center}
{\large\bf Abstract}
\end{center}
\begin{center}
\begin{minipage}{4.55in}

We construct the spectral curve of $\gamma$-deformed $AdS/CFT$ 
from the strong coupling scaling limit of the $T$-system. 
As we interpret the twisted $T$-functions in the classical limit as characters of the highest weight representations of the $psu(2,2|4)$ symmetry group, we compute the twisted quasimomenta which characterize
classical integrability and analyze their  analytic and 
asymptotic properties. These twisted quasimomenta 
are compared to
 Beisert-Roiban Bethe ansatz equations and classical string solutions.


\end{minipage}
\end{center}
\end{titlepage}

\newpage

\section{Introduction}
The integrability approach has been rapidly developed in the
 $AdS/CFT$ correspondence  during the last decade \cite{Beisert:2010jr}. Quantum integrability, as inspired by perturbative integrability \cite{Minahan:2002ve, Bena:2003wd, Beisert:2005bm}, appeared in both gauge and string theory and has led to various exact results such as the $S$-matrix \cite{Beisert:2005tm, Arutyunov:2006yd}, the all-loop Bethe ansatz equations (BAEs) \cite{Beisert:2005fw} and the Thermodynamic Bethe ansatz (TBA) \cite{Gromov:2009tv, Bombardelli:2009ns, Arutyunov:2009ur}. Those have passed many nontrivial tests till now. Among these, TBA is a tool which allows us to determine the exact spectrum including all finite size effects. However, it is  practically difficult to obtain analytical and numerical results by solving directly the TBA equations.\footnote{See \cite{Gromov:2011cx} for recent progress as fundamental solutions.}  

It is remarkable that the TBA can be reformulated in terms of the $T$-system (supplemented with analytic and asymptotic information),  which is mathematically more elegant. 
The $(a,s)$ integer lattice (T-hook) structure of the $T$-system is closely connected to the $psu(2,2|4)$
 symmetry of the model \cite{Gromov:2010vb, Gromov:2010kf}.
Interestingly, solutions of the $T$-system can be obtained by generating functionals which follow from consistencies between different B\"acklund flows \cite{Gromov:2010vb, Kazakov:2007fy}.
Especially, in the asymptotic limit, one can determine the finite form
of these generating functionals as the entire T-hook is decoupled into left and right wings.
Additionally, the classical limit of $T$-functions can be interpreted as the character in the highest weight representation of the symmetry group. Therefore, through the character formula, one can easily compute 
the eigenvalues of the classical transfer matrix in the scaling
(asymptotic and strong coupling) limit and obtain the quasimomenta which characterize the integrability of classical strings. The quasimomenta are also useful to analyze (quasi)classical strings and even exact quantum strings. 

Based on evidences from previous investigations we assume  that the  $T$-system is universal for a given model. In other words, we use the same $T$-system equations and T-hook boundary conditions for models which  differ from each other but share the same bulk symmetry. The only differences are the asymptotic 
behaviour and the analytic structure as we  move from one model to the other. For example, $AdS_5/CFT_4$ models with integrable boundaries and $\beta$ deformation are exactly of this type \cite{Bajnok:2012xc, Bajnok:2013sza, Correa:2012hh, Drukker:2012de, Bajnok:2013wsa, Gromov:2010dy}. 

Among examples of such  universal $T$-systems, the $\beta$ deformed $AdS_5/CFT_4$ is one of the most interesting ones. On the gauge theory side it is a specific marginal deformation of the  ${\cal N}=4$ SYM with real $\beta$ and $h=0 $ in the Leigh-Strassler superpotential \cite{LS95}.\footnote{Recently, it was demonstrated that the complex $\beta$-deformation on the string theory side is generally not integrable \cite{Giataganas:2013dha}.} On the other hand, on the dual string theory side, the deformation could be done by $TsT$-transformations on the $AdS_5 \times S^5$ background \cite{LM05}. On both sides of the deformed duality, integrability seems to be unbroken because in the string theory regime the Lax pair structure is preserved under $TsT$-transformation \cite{F05, Alday:2005ww}, while in the  gauge theory regime the deformation of the
spin-chain $R$-matrix is of the  Drinfeld-Reshetikhin form \cite{BR05}. 
Furthermore, the $\beta$ deformation naturally has a three parameter family of generalizations ($\gamma_{i}$-deformation) on each side of the duality \cite{F05}. In  the $\beta$-deformed and $\gamma_i$-deformed models, the ${\cal N}=4$ supersymmetry is reduced to ${\cal N}=1$ or is totally broken, but conformal symmetry is preserved.\footnote{Very recently, 
non-conformality of the $\gamma_{i}$-deformation is analyzed in \cite{Fokken:2013aea}. Nevertheless, we will consider the $\gamma_i$-deformed case which can be easily reduced to the $\beta$-deformed case with $\gamma_{1,2,3}=\beta$ in the final formulas. However, there still remains many interesting points in the deformation context: see \cite{Fokken:2013mza} for prewrapping phenomena or  \cite{vanTongeren:2013gva} for a nice review.}
In the $\gamma_i$-deformed duality, the all-loop BAEs, the exact $S$-matrix and the TBA equations were developed and it was shown that those are consistent with direct perturbative computations \cite{BR05, Ahn:2010ws, Ahn:2011xq}.  
The general twisted spectral curves were first studied in \cite{Gromov:2007ky}. By setting carefully the twist parameters, they could recover the Beisert-Roiban BAEs. 
Especially, it was found that the undeformed $Y$-system can  
be applied to the deformed model and the weak coupling limit of the $Y$-system is consistent with  perturbative 
computations \cite{Gromov:2010dy}. This shows that the universality of $Y$ and $T$-systems works well in the gauge theory regime of the $\gamma_i$-deformed duality.
In this letter we would like to analyze whether the $T$-system describes properly the results in the strong coupling regime.

This letter is organized as follows. Based on the $T$-system in the deformed models, we consider the strong coupling scaling limit of the  
$\gamma_{i}$-deformed $AdS_5/CFT_4$. Actually, we analyze this limit
 of the twisted generating functionals given in \cite{Gromov:2010dy}. In the procedure, we introduce gauge functions for the $T$-system 
to ensure the correct middle node Bethe equation. Next, 
by  comparing the expansion appearing in the scaling limit to the character of the highest weight representation of $psu(2,2|4)$, we compute the twisted quasimomenta for the $\gamma_{i}$-deformed $AdS/CFT$ and analyze their analytic and asymptotic properties. We also show that the twisted quasimomenta are consistent with the Beisert-Roiban BAEs and check their validity through classical string analysis.

\section{$T$-system and spectral curve of $\gamma_{i}$-deformed $AdS/CFT$}
The exact spectral information of the $AdS/CFT$ correspondence could be computed from the $Y$-functions which are solutions of the $Y$-system equations:
\begin{equation}
Y_{a,s}^{+} Y_{a,s}^{-} = \frac{(1+Y_{a,s+1})(1+Y_{a,s-1})}{(1+1/Y_{a+1,s})(1+1/Y_{a-1,s})}, \quad ; \quad Y^{\pm}(u) \equiv Y(u\pm i/2),
\end{equation}
where the $Y$-functions are defined on an integer lattice called $T$-hook \cite{Gromov:2009tv}.
Then, the exact energy expression of a particle is given by
\begin{equation}
\Delta = \sum_{j} \epsilon_{1}(u_{4,j}) + \sum_{a=1}^{\infty} \int \frac{du}{2\pi i} \frac{\partial \epsilon_a^{{\rm mir}}(u)}{\partial u} \log{(1+Y_{a,0}^{{\rm mir}}(u))},
\end{equation}
where $\epsilon_1 (u) = 1+ 2 i g (1/x^{+}(u) - 1/x^{-}(u))$ is the energy of a fundamental excitation (magnon) in the physical branch and the integral term  is the exact finite size correction.\footnote{$x(u)$ is the function of $u$ such as $x(u)+\frac{1}{x(u)} = \frac{u}{g}$.
} 
If we express the $Y$-functions in terms of the $T$-functions as
\begin{equation}
Y_{a,s}=\frac{T_{a,s+1} T_{a,s-1}}{T_{a+1,s} T_{a-1,s}}, 
\label{YT}
\end{equation}
then the $Y$-system equations can be rewritten into the $T$-system equations as \cite{Gromov:2008gj}
\begin{equation}
T_{a,s}^{+} (u) T_{a,s}^{-} (u) = T_{a+1,s} (u) T_{a-1,s} (u) + T_{a,s+1} (u) T_{a,s-1} u),
\end{equation}
where the $T$-functions (solutions of the $T$-system) can be obtained by generating functionals.

In \cite{Gromov:2010dy}, Gromov and Levkovich-Maslyuk proposed the twisted generating functional in the $\beta$-deformed $AdS_5/CFT_4$ duality and demonstrated that the deformation effect is captured  only by the asymptotic solution with the same $Y$ and $T$-systems. Then, the specific asymptotic solution of the $Y$-system could be expressed in terms of the $T$-functions which are given by the generating functionals ${\cal W}_{R,L}$ : 
\begin{equation} 
{\cal W}_{R,L} = \sum_{s=0}^{\infty} D^s T_{1,s}^{R,L} D^s , \quad
{\cal W}_{R,L}^{-1} = \sum_{a=0}^{\infty} (-1)^{a} D^a T_{a,1}^{R,L} D^a .
\end{equation} 
In the decoupling limit the generating functionals for the left and right wings of the $T$-hook can be written in the following closed forms:
\begin{eqnarray}
{\cal W}_{R} &=& \frac{1}{1-\frac{1}{\tau_{R}}D F_R \frac{B^{(-)-} Q_{1}^{+}}{B^{(+)-} Q_{1}^{-}}D}  \left( 1- D F_R \frac{Q_{1}^{+} Q_{2}^{--} }{Q_{1}^{-} Q_{2} }D \right) \cr
&\times & \left( 1-  D F_R \frac{Q_{2}^{++} Q_{3}^{-} }{Q_{2} Q_{3}^{+}} D  \right) \frac{1}{1- \tau_{R} D F_R \frac{R^{(+)+} Q_{3}^{-}}{R^{(-)+} Q_{3}^{+}}D} \label{wr} \\
{\cal W}_{L} &=& \frac{1}{1-\frac{1}{\tau_{L}}D F_L \frac{B^{(-)-} Q_{7}^{+}}{B^{(+)-} Q_{7}^{-}}D}  \left(1- D F_L \frac{Q_{7}^{+} Q_{6}^{--} }{Q_{7}^{-} Q_{6} }D \right) \cr
&\times & \left( 1-  D F_L \frac{Q_{6}^{++} Q_{5}^{-} }{Q_{6} Q_{5}^{+}} D  \right) \frac{1}{1- \tau_{L} D F_L \frac{R^{(+)+} Q_{5}^{-}}{R^{(-)+} Q_{5}^{+}}D} \label{wl}
\end{eqnarray}
Here, $D$ is the shift operator, $D\equiv e^{-i \partial_{u}/2} $, and $\tau_{R,L}$ are twists for each wing such as\footnote{Twists were introduced for general integrable deformation in \cite{Gromov:2010dy}. Here, we defined $\tau_{R,L}$ as $\tau_{R,L}^1$ in \cite{Gromov:2010dy} and ignored $\tau_{R,L}^2$ as those are irrelevant for the $\gamma_{i}$-deformed theory. } 
\begin{eqnarray}
\tau_{R} &=& e^{-i(K_4 (\delta_1 +\delta_2 +\delta_3)+K_5 (-\delta_1 -\delta_2 -\delta_3)+K_7 (\delta_1 -\delta_2)+K_1 \delta_3 -L \delta_3)},\cr 
\tau_{L} &=& e^{-i(K_3 (\delta_1 +\delta_2 +\delta_3)+K_4 (-\delta_1 -\delta_2 -\delta_3)-K_7 \delta_1 +K_1 (\delta_2 +\delta_3 ) +L \delta_1)}. \label{twi1}
\end{eqnarray}
Also, the functions $R^{(\pm)}(u)$, $B^{(\pm)}(u)$, $Q_{j} (u)$ and the integers $K_j  (j=1, \ldots ,7)$ are defined as
\begin{equation}
R^{(\pm)}(u) =  \prod_{i=1}^{K_4}\left(x(u)-x^{\mp}_{4,i} \right) \,;\
B^{(\pm)}(u) = \prod_{i=1}^{K_4}\left(\frac{1}{x(u)}-x^{\mp}_{4,i} \right) \,;\ Q_j (u) = \prod_{i=1}^{K_j}(u-u_{j,i}), \\
\end{equation}
while $\delta_j (j=1, \dots ,3)$ are constants characterizing the deformation.
In the twisted generating functionals (\ref{wr}), (\ref{wl}), we introduced the functions $F_{R,L}$ by using the gauge degrees of freedom of the $T$-functions as:
\begin{eqnarray}
F_R &=& \left(\frac{x^-}{x^+}\right)^{\frac{L}{2}} \prod\limits_{j=1}^{K_4} \sigma (u, u_{4,j}) \frac{R^{(-)+}}{R^{(+)+}} \sqrt{\frac{Q_4^{++}}{Q_4^{--}}} \frac{B_1^{-} B_3^{+}}{B_1^{+} B_3^{-}}, \\
F_L &=& \left(\frac{x^-}{x^+}\right)^{\frac{L}{2}} \prod\limits_{j=1}^{K_4} \sigma (u, u_{4,j}) \frac{R^{(-)+}}{R^{(+)+}} \sqrt{\frac{Q_4^{++}}{Q_4^{--}}} \frac{B_7^{-} B_5^{+}}{B_7^{+} B_5^{-}},
\end{eqnarray}
where $B_{l}$ is defined as $B_{l}=\prod\limits_{j=1}^{K_l} (1/x(u) -x_{l,j}) $.
Then, the middle node Bethe equation for the momentum carrying root from (\ref{YT}) can be simply written as $Y_{a,0}(u_{4,j})\simeq T_{a,1}^{L}(u_{4,j}) T_{a,1}^{R}(u_{4,j}) = -1$ in the asymptotic limit.

\subsection{Strong coupling scaling limit and twisted quasimomenta}
In the undeformed $AdS_5 \times S^5$ string $\sigma$ model the classical $T$-functions can be represented by characters of the highest weight representations of the corresponding $psu(2,2|4)$ symmetry. 
Because deformation effects are only captured by the asymptotic solutions,
we can still use the character formula  in the deformed theory.\footnote{On the string side, the $TsT$-transformed $AdS_5 \times S^5$ string model is equivalent to the undeformed string model with appropriate twisted boundary conditions. This fact is a clue for the universality of the $T$-system.}
 
In the strong coupling scaling limit we have the following expansion results :
\begin{eqnarray}
&& \frac{R^{(+)+}}{R^{(-)+}} \simeq  f(x) = e^{i G_{4} (x)}, \qquad \,\ \frac{Q_{j}^{+}}{Q_{j}^{-}} \simeq  \frac{Q_{j}^{++}}{Q_{j}} \simeq g_{j}(x) = e^{i \left(H_j (x) + H_j (1/x)  \right)}, \\
&& \frac{B^{(-)-}}{B^{(+)-}} \simeq  {\tilde f}(x) = e^{- i G_{4} (1/x)}, \quad \frac{Q_{j}^{-}}{Q_{j}^{+}} \simeq  \frac{Q_{j}^{--}}{Q_{j}} \simeq {\tilde g}_{j}(x) = e^{-i \left(H_j (x) + H_j (1/x)  \right)}, 
\end{eqnarray}
where $G_4 (x)$ and $H_{j}(x)$ are the resolvents defined as 
\begin{equation}
G_{4}(x) =\sum_{j=1}^{K_4}\frac{x_{4,j}^2}{g(x_{4,j}^2 -1)} \frac{1}{x-x_{4,j}}, \qquad H_{j}(x) =\sum_{k=1}^{K_j}\frac{x^2}{g(x^2 -1)} \frac{1}{x-x_{j,k}}. 
\end{equation}
In this scaling limit, the gauge functions $F_{R,L}$ which were introduced from the gauge degrees of freedom are reduced to
\begin{eqnarray}
F_R &\simeq & {\rm exp}\left[-i \left(\frac{L x/(2 g)- G_4'(0) x}{x^2-1} -H_1(1/x)+H_3 (1/x) \right)\right] \equiv F_R^S, \\
F_L &\simeq & {\rm exp}\left[-i \left(\frac{L x/(2 g)- G_4'(0) x}{x^2-1} -H_7(1/x)+H_5 (1/x) \right)\right] \equiv F_L^S.
\end{eqnarray} 
where $G_4'(x)$ is the derivative of $G_4(x)$ wrt. $x$ and we used the strong coupling expressions of $x^{\pm}$ and $\log{\sigma (u, u_{4,j})}$  as follows:
\begin{eqnarray}
 x^{\pm} &\simeq &  x \pm \frac{i}{2 g} \frac{x^2}{x^2 -1}, \\
\log \sigma(u,u_{4,j}) &\simeq & {\frac{i(x-x_{4,j}) }{g(-1+x^2) (-1+x_{4,j} x) (-1+x_{4,j}^2)} } .
\end{eqnarray}
Meanwhile, since the shift operator $D$ becomes a formal expansion parameter in the scaling limit,
the generating functionals for each wing can be reduced to
\begin{eqnarray}
{\cal W}_{R} &=& \frac{\left(1- F_R^S g_{1} (x) {\tilde g}_{2}(x)  D^2\right) \left( 1- F_R^S g_{2} (x) {\tilde g}_{3}(x)  D^2  \right) }{\left(1- \frac{1}{\tau_R} F_R^S {\tilde f}(x) g_{1}(x) D^2\right) \left(1- \tau_R F_R^S f(x) {\tilde g}_{3} D^2\right)}, \\
{\cal W}_{L} &=& \frac{\left(1- F_L^S g_{7} (x) {\tilde g}_{6}(x)  D^2\right) \left( 1- F_L^S g_{6} (x) {\tilde g}_{5}(x)  D^2  \right) }{\left(1- \frac{1}{\tau_R} F_L^S {\tilde f}(x) g_{7}(x) D^2\right) \left(1- \tau_R F_L^S f(x) {\tilde g}_{5} D^2\right)}.
\end{eqnarray}
The above generating functionals in the scaling limit can be identified with the $psu(2,2|4)$ character formula : \cite{Gromov:2010vb}
\begin{equation}
W = W^R \times W^{L} = \frac{(1-\mu_1 d^L)(1-\mu_2 d^L)}{(1-\lambda_1 d^L)(1-\lambda_2 d^L)} \times \frac{(1-d^R/\mu_3)(1-d^R/\mu_4)}{(1-d^R/\lambda_3)(1-d^R/\lambda_4)}\quad {\rm with} \quad D^2\equiv d^{R,L}.
\end{equation}
Then, as the quasimomenta are defined by the eigenvalues $\lambda_j , \mu_j$ of the classical transfer matrix $T_{a,1}^{R,L}$ such as $\lambda_{j} \equiv e^{-i {\tilde p}_j}, \,\ \mu_{j} \equiv e^{-i {\hat p}_j}$, one can straightforwardly compute the quasimomenta as follows :
\begin{eqnarray}
\hat{p}_1 (x) &=& -\hat{p}_2(1/x) =
+\frac{L x/(2g) - G_4'(0) x}{x^2-1}-H_1 (x) 
+ H_2(x) + H_2(1/x)-H_3 (1/x), \cr
\hat{p}_3 (x) &=& -\hat{p}_4(1/x) =
-\frac{L x/(2g)- G_4'(0) x}{x^2-1}-H_5 (x) 
+ H_6(x)+ H_6(1/x) - H_7(1/x),
\cr
\tilde{p}_1 (x) &=&
+\frac{L x/(2g) + G_4(0)}{x^2-1}-H_1 (x) 
- H_3(1/x) + H_4(1/x)+\phi_{{\tilde 1}} , \cr
\tilde{p}_2 (x) &=& +\frac{L x/(2g) + G_4(0)}{x^2-1}+H_1 (1/x) 
+ H_3(x) - H_4(x)+\phi_{{\tilde 2}}, \cr
\tilde{p}_3 (x) &=&
-\frac{L x/(2g) + G_4(0)}{x^2-1}+H_4 (x) 
- H_5(x)- H_7(1/x)+\phi_{{\tilde 3}}, \cr
\tilde{p}_4 (x) &=& -\frac{L x/(2g) + G_4(0)}{x^2-1}-H_4 (1/x) 
+ H_5(1/x)+ H_7(x)+\phi_{{\tilde 4}},
\label{betaqm}
\end{eqnarray}
where $ \phi_{{\tilde 1}}= -i\log{\tau_{R}} +G_4(0), \quad \phi_{{\tilde 2}}= i\log{\tau_{R}} , \quad \phi_{{\tilde 3}}= -i\log{\tau_{L}} , \quad \phi_{{\tilde 4}} =  i\log{\tau_{L}} -G_4(0).$
Note that we used simple identities between the resolvents such as
\begin{equation}
-\frac{G_4'(0) x}{x^2-1}= -H_4(x)+\frac{G_4(0)}{x^2-1}+G_4 (x)= H_4(1/x)+\frac{G_4(0)}{x^2-1}-G_4 (1/x) +G_4(0).
\end{equation}
Here, $-i\log{\tau_{R,L}}$ can be written as
\begin{eqnarray}
-i\log{\tau_{R}} &=&  \pi((\gamma_3 -\gamma_2)J_1 +(\gamma_1 +\gamma_3)J_2 -(\gamma_1 +\gamma_2)J_3),  \\
-i\log{\tau_{L}} &=&  \pi((\gamma_3 +\gamma_2)J_1 +(\gamma_3 -\gamma_1)J_2 -(\gamma_2 +\gamma_1)J_3),
\end{eqnarray}
because $\delta_{1,2,3}$ of (\ref{twi1}) are written in terms of the deformation parameters $\gamma_{1,2,3}$\footnote{The deformation parameters $\gamma_{1,2,3}$ are introduced in deformed background \cite{F05}. If all $\gamma_{i}$ are the same as $\beta$,  the three parameter (non-supersymmetric) deformed theory reduces to the ${\cal N}=1$ $\beta$-deformed theory. }  as
\begin{equation}
\delta_{1} = -\pi(\gamma_{2} +\gamma_{3}), \quad \delta_{2} = -\pi(\gamma_{1} -\gamma_{2}), \quad \delta_{3} = -\pi(\gamma_{2} -\gamma_{3}). 
\end{equation}
We also used the relation between the numbers of the Bethe roots and the conserved Cartan charges such as\footnote{We choose these relations to make consistent with the results in the twisted transfer matrix formalism \cite{Arutyunov:2010gu}.}
\begin{eqnarray}
&& L \rightarrow J+K_4 + (K_1 -K_3 -K_5 +K_7)/2 , \,\  \quad J \rightarrow  J_1 , \label{rtcar}  \\
&& K_1 + K_3 + K_5 +K_7 -2 K_4 \rightarrow -2 J_2 , \qquad \quad
 K_1 + K_3 - K_5 -K_7  \rightarrow -2 J_3. \nonumber
\end{eqnarray}
Furthermore, we can impose the momentum quantization condition for the $\gamma_i$-deformed $AdS/CFT$ as $-G_4 (0)$ is identified with the 1st conserved charge $p$, which is the momentum.
\begin{equation}
p=-i \sum_{j=1}^{K_4} \log{\frac{x_{4,j}^{+}}{x_{4,j}^{+}}} = 2\pi (J_2 \gamma_3 - J_3 \gamma_2 ) \label{momcon}
\end{equation}
Then, we finally have the following expressions:
\begin{eqnarray}
\phi_{{\tilde 1}} &=& +\pi((\gamma_3 -\gamma_2)J_1 +(\gamma_1 -\gamma_3)J_2 +(\gamma_2 -\gamma_1)J_3), \cr
\phi_{{\tilde 2}} &=& -\pi((\gamma_3 -\gamma_2)J_1 +(\gamma_1 +\gamma_3)J_2 -(\gamma_1 +\gamma_2)J_3), \cr
 \phi_{{\tilde 3}} &=& +\pi((\gamma_3 +\gamma_2)J_1 +(\gamma_3 -\gamma_1)J_2 -(\gamma_2 +\gamma_1)J_3), \cr 
 \phi_{{\tilde 4}} &=& -\pi((\gamma_3 +\gamma_2)J_1 -(\gamma_1 +\gamma_3)J_2 +(\gamma_2 -\gamma_1)J_3). \label{deform}
\end{eqnarray}
These twists are closely related to the twisted boundary conditions (\ref{TBC}) of the isometry angles of $S^5$. Note that the twisted boundary conditions (\ref{TBC}) can be understood as the twists of the worldsheet fields $Z_{\alpha {\dot \alpha}}, Y_{a {\dot a}}, \theta_{a {\dot \alpha}}, \eta_{{\dot a} \alpha}$ of the light-cone gauge string theory \cite{Arutyunov:2010gu, deLeeuw:2012hp}. In the $\gamma_i$-deformed case, we just have $\alpha_l =\pi (J_1 (\gamma_3 -\gamma_2)+(J_2 -J_3)\gamma_1)$ and $\alpha_r =\pi ((J_1 (\gamma_2 +\gamma_3 ) - (J_2 +J_3 ) \gamma_1 )$ \cite{vanTongeren:2013gva}. Then, we get the following relations between $\alpha_{l,r}$ of \cite{vanTongeren:2013gva} and the twists of the quasimomenta:
\begin{equation}
\phi_{{\tilde 1}} = -\frac{p}{2} +\alpha_l ,\quad \phi_{{\tilde 2}} = -\frac{p}{2} -\alpha_l ,\quad
\phi_{{\tilde 3}} = +\frac{p}{2} +\alpha_r ,\quad \phi_{{\tilde 1}} = +\frac{p}{2} -\alpha_r .
\end{equation}
On the other hand, the twists of the quasimomenta (\ref{deform}) can be also expressed by using the twists of the isometry angles (\ref{TBC}) as follows  : 
\begin{eqnarray}
\phi_{{\tilde 1}} &=& +(\delta\phi_{1}+\delta\phi_{2}+\delta\phi_{3}),\qquad \phi_{{\tilde 2}} = +(\delta\phi_{1}-\delta\phi_{2}-\delta\phi_{3}) ,\cr
\phi_{{\tilde 3}} &=& -(\delta\phi_{1}-\delta\phi_{2}+\delta\phi_{3}) ,\qquad \phi_{{\tilde 1}} = -(\delta\phi_{1}+\delta\phi_{2}-\delta\phi_{3}). 
\end{eqnarray}
We finally notice that these results are consistent with the eigenvalues $\Lambda$ of the twisted monodromy matrix in \cite{Alday:2005ww}.

\subsection{Properties of twisted quasimomenta}
In this subsection we analyze the analytic and asymptotic properties of the twisted quasimomenta (\ref{betaqm}).
Firstly, the synchronization of the residues at $x=\pm 1$ is automatically satisfied as in the deformed theory  the twist parameters are just constants. 

The inversion symmetries between ${\hat p}_{i}$ are simply given as 
\begin{equation}
\hat{p}_{1,2}(1/x)=-\hat{p}_{2,1}(x),\quad 
\hat{p}_{3,4}(1/x)=-\hat{p}_{4,3}(x),
\end{equation}
because the $AdS$ part does not  change in the $\gamma_i$-deformation.
On the other hand, the inversion symmetries between ${\tilde p}_{i}$ have nontrivial deformation effects which came from $G_4 (0)$:
\begin{equation}
\tilde{p}_{1,2}(1/x)=-\tilde{p}_{2,1}(x)-2\pi (J_2 \gamma_3 -J_3 \gamma_2)\, ;\quad 
\tilde{p}_{3,4}(1/x)=-\tilde{p}_{4,3}(x)+2\pi (J_2 \gamma_3 -J_3 \gamma_2)
\end{equation}

Also, one can analyze the large $x$ asymptotics from the twisted quasimomenta:
\begin{eqnarray}
{\hat p}_{1}(x) & \simeq & \frac{2\pi}{\sqrt{\lambda} x} (\Delta -S_1 +S_2), \qquad \qquad
{\hat p}_{2}(x) \simeq  \frac{2\pi}{\sqrt{\lambda} x} (\Delta +S_1 -S_2), \cr
{\hat p}_{3}(x) & \simeq & \frac{-2\pi}{\sqrt{\lambda} x} (\Delta +S_1 +S_2), \qquad \qquad
{\hat p}_{4}(x) \simeq  \frac{-2\pi}{\sqrt{\lambda} x} (\Delta -S_1 -S_2),
\cr
{\tilde p}_{1}(x) & \simeq & \frac{2\pi}{\sqrt{\lambda} x} (J_1 +J_2 -J_3) +\phi_{{\tilde 1}}, \qquad
{\tilde p}_{2}(x)  \simeq  \frac{2\pi}{\sqrt{\lambda} x} (J_1 -J_2 +J_3) +\phi_{{\tilde 2}},  \cr
{\tilde p}_{3}(x) & \simeq & \frac{-2\pi}{\sqrt{\lambda} x} (J_1 -J_2 -J_3) +\phi_{{\tilde 3}}, \qquad 
{\tilde p}_{4}(x)  \simeq  \frac{-2\pi}{\sqrt{\lambda} x} (J_1 +J_2 +J_3)+\phi_{{\tilde 4}}. \label{largexasymp}
\end{eqnarray} 

Let us comment on the gradings. We used the $su(2)$ grading as in \cite{Gromov:2010dy}. If we repeat the same computations from the twisted generating functionals in the $sl(2)$ grading, we could obtain the twisted quasimomenta in the $sl(2)$ grading. Those quasimomenta in different gradings are connected to each other through  
the duality transformation.

\subsection{Comparison to BR BAEs} 
The all-loop Bethe ansatz equations for the general integrable deformation were constructed by Beisert and Roiban (BR) \cite{BR05}.
The explicit form of the deformation could be expressed by the product of a phase matrix ${\bold A}$ and a column vector ${\bold K}$ for the excitation numbers.\footnote{For the full expressions, see (5.39) of \cite{BR05} and (6.2) of \cite{Ahn:2010ws}.  }
For the $\gamma_{i}$-deformed $AdS/CFT$,  the ${\bold A}{\bold K}$ column matrix is written as follows\footnote{We used the relation (\ref{rtcar}) in the ${\bold A}{\bold K}$ column matrix. Also, one can easily check that the ${\bold A}{\bold K}$-matrix elements are matched well with the twist factors of the three-spin Bethe equations in \cite{Frolov:2005iq}} :
\begin{equation}
{\bold A}{\bold K}=\left(\begin{array}{c}
-2\pi (J_2 \gamma_3 - J_3 \gamma_2 )\\
-\pi((\gamma_3 -\gamma_2)J_1 +(\gamma_1 -\gamma_3)J_2 +(\gamma_2 -\gamma_1)J_3)\\
0\\
-\pi((\gamma_3 -\gamma_2)J_1 +(\gamma_1 +\gamma_3)J_2 -(\gamma_1 +\gamma_2)J_3)\\
-2\pi((\gamma_1 + \gamma_2 )J_3 -(J_1 +J_2 )\gamma_3 )\\
-\pi((\gamma_3 +\gamma_2)J_1 +(\gamma_3 -\gamma_1)J_2 -(\gamma_2 +\gamma_1)J_3)\\
0\\
-\pi((\gamma_3 +\gamma_2)J_1 -(\gamma_1 +\gamma_3)J_2 +(\gamma_2 -\gamma_1)J_3)\\
\end{array}\right).
\end{equation}
For consistency, the specific differences between the twisted quasimomenta should be matched with the scaling limit of the BR Bethe equations.\footnote{In \cite{Ahn:2012hsa}, twisted quasimomenta were obtained from the scaling limit of the BR BAEs and the giant magnon in the $\beta$-deformed case was studied from the quasimomenta.} These can be checked simply by the following relations :
\begin{eqnarray}
{\bold A}{\bold K}_0 &=& -p, \qquad \quad {\bold A}{\bold K}_1 = -\phi_{{\tilde 1}}, \quad {\bold A}{\bold K}_2 = 0, \quad
{\bold A}{\bold K}_3 = \phi_{{\tilde 2}}, \cr
{\bold A}{\bold K}_4 &=& \phi_{{\tilde 3}}-\phi_{{\tilde 2}}, \quad
{\bold A}{\bold K}_5 = -\phi_{{\tilde 3}}, \quad
{\bold A}{\bold K}_6 = 0, \quad
{\bold A}{\bold K}_7 = \phi_{{\tilde 4}},
\end{eqnarray}
since the elements of ${\bold A}{\bold K}$ appear in the scaled BR Bethe equations separately such as
\begin{eqnarray}
{\hat p}_1 -{\tilde p}_1  &\rightarrow & {\bold A}{\bold K}_1 , \quad
{\hat p}_2 -{\hat p}_1  \rightarrow {\bold A}{\bold K}_2 , \quad
{\tilde p}_2 -{\hat p}_2 \rightarrow  {\bold A}{\bold K}_3 , \cr
{\tilde p}_3 - {\tilde p}_2  &\rightarrow & {\bold A}{\bold K}_4 , \quad
{\hat p}_3 -{\tilde p}_3 \rightarrow {\bold A}{\bold K}_5 , \quad
{\hat p}_4 -{\hat p}_3 \rightarrow {\bold A}{\bold K}_6 , \cr
{\tilde p}_4 - {\hat p}_4  &\rightarrow & {\bold A}{\bold K}_7 . \nonumber
\end{eqnarray}

\subsection{Classical string analysis} 
As a simple check for the twisted quasimomenta, let us describe the rigid circular strings in Lunin-Maldacena background with $\gamma_i =\beta$:
\begin{eqnarray}
ds^{2}_{string}/R^{2} &=& ds^{2}_{AdS_{5}} + \sum_{i=1}^{3}\left(d\rho^{2}_{i}+G\rho^{2}_{i}d\phi^{2}_{i}\right)+G \rho^{2}_{1}\rho^{2}_{2}\rho^{2}_{3}[(\sum_{i=1}^{3}\hat{\beta} d\phi_{i})]^2 \cr
B_{2} &=& R^{2} \hat{\beta} G \left(\rho_1^2 \rho_2^2 d\phi_1 \wedge d\phi_2
                + \rho_2^2 \rho_3^2 d\phi_2 \wedge d\phi_3
                + \rho_3^2 \rho_1^2 d\phi_3 \wedge d\phi_1\right)\cr
G^{-1} &=& 1 +  \hat{\beta}^{2} (\rho_1^2 \rho_2^2 + \rho_2^2 \rho_3^2 + \rho_3^2 \rho_1^2 ) \nonumber
\end{eqnarray}
where the deformation parameter ${\hat \beta}$ is defined as ${\hat \beta}=\sqrt{\lambda} \beta$.
In \cite{F05, deLeeuw:2012hp}, it was shown that any string configuration on Lunin-Maldacena background is equivalent to a string configuration on the undeformed $AdS_5 \times S^5$ with the following twisted boundary conditions \footnote{See \cite{Ahn:2010ws, Ahn:2012hs} for different but equivalent integrable formulations. }:
\begin{equation}
\delta\phi_i \equiv \phi_i(2\pi)-\phi_i(0)=2 \pi (n_i - \epsilon_{ijk} \gamma_j J_k),\,\,(i,j,k=1,2,3). \label{TBC}
\end{equation}
One can notice that the momentum $p$ in the light-cone gauge strings which is identified with $\delta\phi_1$ is exactly matched with the cyclic condition (\ref{momcon}) in the Bethe equations.  

Then, let us consider the three spin rotating string ans\"atze such as
\begin{equation}
t=\kappa \tau,\,\, \theta=\theta_{0},\,\, \alpha=\alpha_{0},\,\,  \phi_{i}=\omega_{i} \tau + m_{i} \sigma,\,\ (i=1,2,3).
\end{equation}
These ans\"atze on  Lunin-Maldacena background satisfy the Virasoro constraints and the equations of motions. On the other hand, one can clearly check that the twisted boundary conditions (\ref{TBC}) in the rigid circular strings are nothing but shifts of the mode numbers  as
\begin{equation}
m_1 \rightarrow m_1 - 2\pi\beta(J_3 -J_2),\,\,\, m_2 \rightarrow m_2 - 2\pi\beta (J_1 - J_3 ),\,\,\, m_3 \rightarrow m_3 - 2\pi\beta (J_2 -J_1 ). \nonumber
\end{equation}
In other words, if we consider the same ans\"atze in the undeformed $AdS_5 \times S^5$ with (\ref{TBC}), the resulting Virasoro constraints and the equations of motion are exactly the same as those in Lunin-Maldacena background.

Especially, all the constraints are simplified in case of the $su(2)$ rigid circular string with $m_1 = - m_2 =m$ and $J_1 = J_2 = J/2$ because of $m_3 =0 =J_3$. 
Let us consider the quasimometa of circular strings on the undeformed $S^5$ \cite{Gromov:2007aq} such as 
\begin{eqnarray}
{\tilde p}_1 &=& \frac{2\pi x}{\sqrt{\lambda} (x^2 -1)} \sqrt{\lambda m^2/x^2 + J^2}, \quad
{\tilde p}_2 = \frac{2\pi x}{\sqrt{\lambda} (x^2 -1)} \sqrt{\lambda m^2 x^2 + J^2} -m, \cr
{\tilde p}_3 &=& \frac{-2\pi x}{\sqrt{\lambda} (x^2 -1)} \sqrt{\lambda m^2 x^2 + J^2}+m, \quad
{\tilde p}_4 = \frac{-2\pi x}{\sqrt{\lambda} (x^2 -1)} \sqrt{\lambda m^2/x^2 + J^2}. \nonumber
\end{eqnarray}
The quasimomenta for the $AdS_5$ are just given as ${\hat p}_{1,2}(x) = -{\hat p}_{3,4}(x) = \frac{2\pi \Delta x}{\sqrt{\lambda} (x^2 -1)}$. \\

We can obtain the twisted quasimomenta from the untwisted quasimomenta by substituting $m \rightarrow m+ \pi \beta J$. Then, we get the extra twist factors $(0, -\pi \beta J, +\pi \beta J, 0)$ in ${\tilde p}_{1,2,3,4}$ with the shifted $m$ in the square roots of the quasimomenta.
These are really consistent with (\ref{deform}) for $J_1 = J_2 = J/2$ and $J_3 =0$. Therefore, the classical dispersion relation of the $su(2)$ rigid circular string on Lunin-Maldacena background can be written as \cite{Frolov:2005ty} 
\begin{equation}
\Delta = \sqrt{J^2 + \lambda (m+ \pi \beta J)^2}.
\end{equation}

\section{Concluding remarks}
In this letter we have analyzed the strong coupling scaling limit of the $\gamma_{i}$-deformed $AdS/CFT$. We have assumed that one can still use the $psu(2,2|4)$ character formula in the deformed case and have obtained the twisted quasimomenta by comparing the character formula to the scaling limit of the twisted generating functionals. 
We have also analyzed the analytic properties of the twisted quasimomenta.
Finally, we have shown that the twisted quasimomenta for the $\gamma_{i}$-deformation are consistent with both of the Beisert-Roiban BAEs and classical string analysis.

The integrability of the $\gamma_i$-deformed theory is mainly based on the fact that the deformation effects can be treated separately from the bulk symmetry or the $T$-system. 
On the gauge theory side, the deformed spin-chain Hamiltonian through the Drinfeld-Reshetikhen twists of $R$-matrix can be replaced to undeformed Hamiltonian by specific changes of basis \cite{Berenstein:2004ys}. On the string theory side, the deformed background through $TsT$-transformation can be replaced to undeformed $AdS_5 \times S^5$ by twisted boundary conditions of the isometry angles of  $S^5$. In \cite{F05, Alday:2005ww}, the quasimomenta are studied by considering the twisted Lax connections through the twisted boundary conditions. On the other hand, we used the universal $T$-system with the asymptotic solutions for the deformed theory and could check that our twisted quasimomenta are consistent with direct results from the twisted monodromy matrix of the $TsT$-transformed $AdS_5 \times S^5$. 

Recently, it was argued that the quasimomenta which characterize the classical spectral information are needed to read off the asymptotic information of the quantum spectral curve through the $P-\mu$ system \cite{Gromov:2013pga, Gromov:2013qga} in the form 
\begin{equation}
P_{a} \simeq e^{\int {\tilde p}_a(u) du} \nonumber.
\end{equation}
Therefore, the twisted quasimomenta for the $\gamma_{i}$-deformation would be helpful when we would like to construct the $P-\mu$ system for the $\gamma_{i}$-deformed theory. This would be definitely one of the most important directions for future research.

In this letter, we assumed the real deformation parameters in the $\gamma_i$-deformed theory. On the other hand, the complex $\beta$-deformed theory is known to be generally not integrable \cite{Giataganas:2013dha}. It would be interesting to understand the non-integrability in the $T$-system level.

There exists some other integrable deformed models where the techniques in this paper can be applied. For example, the marginally deformed model of ABJM \cite{Imeroni:2008cr, He:2013hxd} and the more interesting quantum deformation called $Q$-deformation \cite{Arutyunov:2012zt, Arutyunov:2012ai, Delduc:2013qra, Arutyunov:2013ega} were known as preserving integrability, even though the exact $AdS/CFT$ pictures are still unclear. Especially, as the generating functionals are known in case of the $Q$-deformation \cite{Arutyunov:2012ai}, it would be interesting to construct the spectral curves through their classical strong coupling limit and compare those to direct constructions from the lax connections in \cite{Delduc:2013qra}.


\section*{Acknowledgements}
I thank Zoltan Bajnok and Laszlo Palla for useful comments on the manuscript. 
This work was supported by the Hungarian scholarship (type D) through the Balassi Institute and by a Lend\"ulet grant.


\begin{thebibliography}{}


\bibitem{Beisert:2010jr}
  N.~Beisert, C.~Ahn, L.~F.~Alday, Z.~Bajnok, J.~M.~Drummond, L.~Freyhult, N.~Gromov and R.~A.~Janik {\it et al.},
  ``Review of AdS/CFT Integrability: An Overview,''  Lett.\ Math.\ Phys.\  {\bf 99}, 3 (2012)  [arXiv:1012.3982 [hep-th]].


\bibitem{Minahan:2002ve}
  J.~A.~Minahan and K.~Zarembo,
  ``The Bethe ansatz for $N=4$ superYang-Mills,''
  JHEP {\bf 0303} (2003) 013
  [hep-th/0212208].


\bibitem{Bena:2003wd}
  I.~Bena, J.~Polchinski and R.~Roiban,
  ``Hidden symmetries of the $AdS_5 \times S^5$ superstring,''
  Phys.\ Rev.\ D {\bf 69} (2004) 046002
  [hep-th/0305116].

\bibitem{Beisert:2005bm}
  N.~Beisert, V.~A.~Kazakov, K.~Sakai and K.~Zarembo,
  ``The Algebraic curve of classical superstrings on $AdS_5 \times S^5$,''
  Commun.\ Math.\ Phys.\  {\bf 263} (2006) 659
  [hep-th/0502226].


\bibitem{Beisert:2005tm}
  N.~Beisert,
  ``The $SU(2|2)$ dynamic S-matrix,''
  Adv.\ Theor.\ Math.\ Phys.\  {\bf 12}, 945 (2008)
  [arXiv:hep-th/0511082].



\bibitem{Arutyunov:2006yd}
  G.~Arutyunov, S.~Frolov and M.~Zamaklar,
  ``The Zamolodchikov-Faddeev algebra for $AdS_{5}\times S^{5}$ superstring,''
  JHEP {\bf 0704}, 002 (2007)
  [arXiv:hep-th/0612229].



\bibitem{Beisert:2005fw}
  N.~Beisert and M.~Staudacher,
  ``Long-range $psu(2,2|4)$ Bethe Ansatze for gauge theory and strings,''
  Nucl.\ Phys.\ B {\bf 727} (2005) 1
  [hep-th/0504190].
  
\bibitem{Gromov:2009tv}
  N.~Gromov, V.~Kazakov and P.~Vieira,
  ``Exact Spectrum of Anomalous Dimensions of Planar $N=4$ Supersymmetric Yang-Mills Theory,''
  Phys.\ Rev.\ Lett.\  {\bf 103} (2009) 131601
  [arXiv:0901.3753 [hep-th]].
  
\bibitem{Bombardelli:2009ns}
  D.~Bombardelli, D.~Fioravanti and R.~Tateo,
  ``Thermodynamic Bethe Ansatz for planar $AdS/CFT$: A Proposal,''
  J.\ Phys.\ A {\bf 42} (2009) 375401
  [arXiv:0902.3930 [hep-th]].
  
\bibitem{Arutyunov:2009ur}
  G.~Arutyunov and S.~Frolov,
  ``Thermodynamic Bethe Ansatz for the $AdS_5 \times S^5$ Mirror Model,''
  JHEP {\bf 0905} (2009) 068
  [arXiv:0903.0141 [hep-th]].    



\bibitem{Gromov:2011cx}
  N.~Gromov, V.~Kazakov, S.~Leurent and D.~Volin,
  ``Solving the $AdS/CFT$ Y-system,''
  JHEP {\bf 1207} (2012) 023
  [arXiv:1110.0562 [hep-th]].


\bibitem{Gromov:2010vb}
  N.~Gromov, V.~Kazakov and Z.~Tsuboi,
  ``$PSU(2,2|4)$ Character of Quasiclassical AdS/CFT,''
  JHEP {\bf 1007} (2010) 097
  [arXiv:1002.3981 [hep-th]].

\bibitem{Gromov:2010kf}
  N.~Gromov and V.~Kazakov,
  ``Review of AdS/CFT Integrability, Chapter III.7: Hirota Dynamics for Quantum Integrability,''
  Lett.\ Math.\ Phys.\  {\bf 99} (2012) 321
  [arXiv:1012.3996 [hep-th]].


\bibitem{Kazakov:2007fy}
  V.~Kazakov, A.~S.~Sorin and A.~Zabrodin,
  ``Supersymmetric Bethe ansatz and Baxter equations from discrete Hirota dynamics,''
  Nucl.\ Phys.\ B {\bf 790} (2008) 345
  [hep-th/0703147 [HEP-TH]].


\bibitem{Bajnok:2012xc}
  Z.~Bajnok, R.~I.~Nepomechie, L.~Palla and R.~Suzuki,
  ``Y-system for $Y=0$ brane in planar AdS/CFT,''
  JHEP {\bf 1208} (2012) 149
  [arXiv:1205.2060 [hep-th]].

\bibitem{Bajnok:2013sza}
  Z.~Bajnok, M.~Kim and L.~Palla,
  ``Spectral curve for open strings attached to the $Y=0$ brane,''
  arXiv:1311.7280 [hep-th].

\bibitem{Correa:2012hh}
  D.~Correa, J.~Maldacena and A.~Sever,
  ``The quark anti-quark potential and the cusp anomalous dimension from a TBA equation,''
  JHEP {\bf 1208} (2012) 134
  [arXiv:1203.1913 [hep-th]].

\bibitem{Drukker:2012de}
  N.~Drukker,
  ``Integrable Wilson loops,''
  JHEP {\bf 1310} (2013) 135
  [arXiv:1203.1617 [hep-th]].

\bibitem{Bajnok:2013wsa}
  Z.~Bajnok, N.~Drukker, A.~Hegedus, R.~I.~Nepomechie, L.~Palla, C.~Sieg and R.~Suzuki,
  ``The spectrum of tachyons in AdS/CFT,''
  arXiv:1312.3900 [hep-th].

\bibitem{Gromov:2010dy}
  N.~Gromov and F.~Levkovich-Maslyuk,
  ``Y-system and $\beta$-deformed N=4 Super-Yang-Mills,''
  J.\ Phys.\ A {\bf 44} (2011) 015402
  [arXiv:1006.5438 [hep-th]].


\bibitem{LS95} R. G. Leigh and M. J. Strassler,
``Exactly marginal operators and duality in four-dimensional
$\mathcal{N}=1$ supersymmetric gauge theory'', Nucl. Phys. {\bf B
447} 95 (1995), [arXiv:hep-th/9503121].

\bibitem{Giataganas:2013dha}
  D.~Giataganas, L.~A.~Pando Zayas and K.~Zoubos,
  ``On Marginal Deformations and Non-Integrability,''
  arXiv:1311.3241 [hep-th].

\bibitem{LM05} O. Lunin and J. Maldacena,
``Deforming field theories with $U(1)\times U(1)$ global symmetry
and their gravity duals'', JHEP {\bf 0505} 033 (2005),
[arXiv:hep-th/0502086].


\bibitem{F05} S. Frolov, ``Lax pair for strings in Lunin-Maldacena background'',
JHEP {\bf 0505} 069 (2005), [arXiv:hep-th/0503201].

\bibitem{Alday:2005ww}
  L.~F.~Alday, G.~Arutyunov and S.~Frolov,
  ``Green-Schwarz strings in TsT-transformed backgrounds,''
  JHEP {\bf 0606} (2006) 018
  [hep-th/0512253].


\bibitem{BR05} N. Beisert and R. Roiban,
``Beauty and the twist: the Bethe ansatz for twisted $\mathcal{N}=4$
SYM'', JHEP {\bf 0508} 039 (2005), [arXiv:hep-th/0505187].


\bibitem{Fokken:2013aea}
  J.~Fokken, C.~Sieg and M.~Wilhelm,
  ``Non-conformality of $\gamma_i$-deformed $N=4$ SYM theory,''
  arXiv:1308.4420 [hep-th].


\bibitem{Fokken:2013mza}
  J.~Fokken, C.~Sieg and M.~Wilhelm,
  ``The complete one-loop dilatation operator of planar real beta-deformed $N=4$ SYM theory,''
  arXiv:1312.2959 [hep-th].


\bibitem{vanTongeren:2013gva}
  S.~J.~van Tongeren,
  ``Integrability of the $AdS_5 \times S^5$ superstring and its deformations,''
  arXiv:1310.4854 [hep-th].


\bibitem{Ahn:2010ws}
  C.~Ahn, Z.~Bajnok, D.~Bombardelli and R.~I.~Nepomechie,
  ``Twisted Bethe equations from a twisted S-matrix,''  JHEP {\bf 1102}, 027 (2011)  [arXiv:1010.3229 [hep-th]].

\bibitem{Ahn:2011xq}
  C.~Ahn, Z.~Bajnok, D.~Bombardelli and R.~I.~Nepomechie,
  ``TBA, NLO Luscher correction, and double wrapping in twisted AdS/CFT,''
  JHEP {\bf 1112} (2011) 059
  [arXiv:1108.4914 [hep-th]].

\bibitem{Gromov:2007ky}
  N.~Gromov and P.~Vieira,
  ``Complete 1-loop test of $AdS/CFT$,''
  JHEP {\bf 0804} (2008) 046
  [arXiv:0709.3487 [hep-th]].

\bibitem{Gromov:2008gj}
  N.~Gromov, V.~Kazakov and P.~Vieira,
  ``Finite Volume Spectrum of 2D Field Theories from Hirota Dynamics,''
  JHEP {\bf 0912} (2009) 060
  [arXiv:0812.5091 [hep-th]].


\bibitem{Arutyunov:2010gu}
  G.~Arutyunov, M.~de Leeuw and S.~J.~van Tongeren,
  ``Twisting the Mirror TBA,''
  JHEP {\bf 1102} (2011) 025
  [arXiv:1009.4118 [hep-th]].


\bibitem{deLeeuw:2012hp}
  M.~de Leeuw and S.~J.~van Tongeren,
  ``The spectral problem for strings on twisted $AdS_5 \times S^5$,''
  Nucl.\ Phys.\ B {\bf 860} (2012) 339
  [arXiv:1201.1451 [hep-th]].



\bibitem{Frolov:2005iq}
  S.~A.~Frolov, R.~Roiban and A.~A.~Tseytlin,
  ``Gauge-string duality for (non)supersymmetric deformations of $N=4$ super Yang-Mills theory,''
  Nucl.\ Phys.\ B {\bf 731} (2005) 1
  [hep-th/0507021].


\bibitem{Ahn:2012hsa}
  C.~Ahn, D.~Bombardelli and M.~Kim,
  ``Finite-size effects of $\beta$-deformed $AdS_5/CFT_4$ at strong coupling,''
  Phys.\ Lett.\ B {\bf 710} (2012) 467
  [arXiv:1201.2635 [hep-th]].




\bibitem{Ahn:2012hs}
  C.~Ahn, M.~Kim and B.~-H.~Lee,
  ``Worldsheet S-matrix of beta-deformed SYM,''
  Phys.\ Lett.\ B {\bf 719} (2013) 458
  [arXiv:1211.4506 [hep-th]].



\bibitem{Gromov:2007aq}
  N.~Gromov and P.~Vieira,
  ``The AdS(5) x S**5 superstring quantum spectrum from the algebraic curve,''
  Nucl.\ Phys.\ B {\bf 789} (2008) 175
  [hep-th/0703191 [HEP-TH]].

\bibitem{Frolov:2005ty}
  S.~A.~Frolov, R.~Roiban and A.~A.~Tseytlin,
  ``Gauge-string duality for superconformal deformations of $N=4$ super Yang-Mills theory,''
  JHEP {\bf 0507} (2005) 045
  [hep-th/0503192].

\bibitem{Berenstein:2004ys}
  D.~Berenstein and S.~A.~Cherkis,
  ``Deformations of $N=4$ SYM and integrable spin chain models,''
  Nucl.\ Phys.\ B {\bf 702} (2004) 49
  [hep-th/0405215].


\bibitem{Gromov:2013pga}
  N.~Gromov, V.~Kazakov, S.~Leurent and D.~Volin,
  ``Quantum spectral curve for $AdS_5/CFT_4$,''
  arXiv:1305.1939 [hep-th].
 

\bibitem{Gromov:2013qga}
  N.~Gromov, F.~Levkovich-Maslyuk and G.~Sizov,
  ``Analytic Solution of Bremsstrahlung TBA II: Turning on the Sphere Angle,''
  JHEP {\bf 1310} (2013) 036
  [arXiv:1305.1944 [hep-th]].

\bibitem{Imeroni:2008cr}
  E.~Imeroni,
  ``On deformed gauge theories and their string/M-theory duals,''
  JHEP {\bf 0810} (2008) 026
  [arXiv:0808.1271 [hep-th]].

\bibitem{He:2013hxd}
  S.~He and J.~-B.~Wu,
  ``Note on Integrability of Marginally Deformed ABJ(M) Theories,''
  JHEP {\bf 1304} (2013) 012
  [arXiv:1302.2208 [hep-th]].

\bibitem{Arutyunov:2012zt}
  G.~Arutyunov, M.~de Leeuw and S.~J.~van Tongeren,
  ``The Quantum Deformed Mirror TBA I,''
  JHEP {\bf 1210} (2012) 090
  [arXiv:1208.3478 [hep-th]].

\bibitem{Arutyunov:2012ai}
  G.~Arutyunov, M.~de Leeuw and S.~J.~van Tongeren,
  ``The Quantum Deformed Mirror TBA II,''
  JHEP
   [JHEP {\bf 1302} (2013) 012]
  [arXiv:1210.8185 [hep-th]].

\bibitem{Delduc:2013qra}
  F.~Delduc, M.~Magro and B.~Vicedo,
  ``An integrable deformation of the $AdS_5 x S^5$ superstring action,''
  arXiv:1309.5850 [hep-th].

\bibitem{Arutyunov:2013ega}
  G.~Arutyunov, R.~Borsato and S.~Frolov,
  ``S-matrix for strings on $\eta$-deformed $AdS_5 \times S^5$,''
  arXiv:1312.3542 [hep-th].
  
\end{thebibliography}
\end{document}